# Buoyant crystals halt the cooling of white dwarf stars

Antoine Bédard[1*†], Simon Blouin[2*†], and Sihao Cheng[3]

[1] Department of Physics, University of Warwick, Coventry, CV4 7AL, UK
[2] Department of Physics and Astronomy, University of Victoria, Victoria, BC V8W 2Y2, Canada
[3] Institute for Advanced Study, Princeton, NJ 08540, USA

* Corresponding authors (antoine.bedard@warwick.ac.uk, sblouin@uvic.ca)
† These authors contributed equally to this work and are joint first authors. Listed order was decided by a quantum coin flip.

**White dwarfs are stellar remnants devoid of a nuclear energy source, gradually cooling over billions of years[1,2] and eventually freezing into a solid state from the inside out[3,4]. Recently, it was discovered that a population of freezing white dwarfs maintains a constant luminosity for a duration comparable to the age of the universe[5], signaling the presence of a powerful yet unknown energy source that inhibits the cooling. For certain core compositions, the freezing process is predicted to trigger a solid–liquid distillation mechanism, due to the solid phase being depleted in heavy impurities[6-8]. The crystals thus formed are buoyant and float up, thereby displacing heavier liquid downward and releasing gravitational energy. Here we show that distillation interrupts the cooling for billions of years and explains all the observational properties of the unusual delayed population. With a steady luminosity surpassing that of some main-sequence stars, these white dwarfs defy their conventional portrayal as dead stars. Our results highlight the existence of peculiar merger remnants[9,10] and have profound implications for the use of white dwarfs in dating stellar populations[11,12].**

The population of delayed white dwarfs was discovered thanks to its clear signature in the *Gaia* Hertzsprung–Russell (H–R) diagram, where it forms an overdensity known as the *Q* branch (Fig. 1)[13]. This overdensity coincides with the predicted location of high-mass white dwarfs undergoing core crystallization[4]. Many *Q*-branch objects have large transverse velocities indicative of old dynamical ages, leading to the conclusion that 5–9% of high-mass white dwarfs stop cooling for at least 8 Gyr[5]. In the canonical crystallization scenario, a stable solid continuously grows from the inside out (Fig. 2a), which releases a modest amount of latent heat and gravitational energy through the partial separation of carbon and oxygen[14-16]. These energy sources slow down the cooling process, but their combined effect is too weak and fails to explain the highly peaked stellar overdensity[17-19] and the long inferred delay time. These observational constraints require that the missing energy source manifests itself only in a narrow temperature range (the surface temperature changing by just 10–15% over the width of the *Q* branch). Although several interpretations have been proposed[5,18-20], none of them is consistent with both observational aspects.

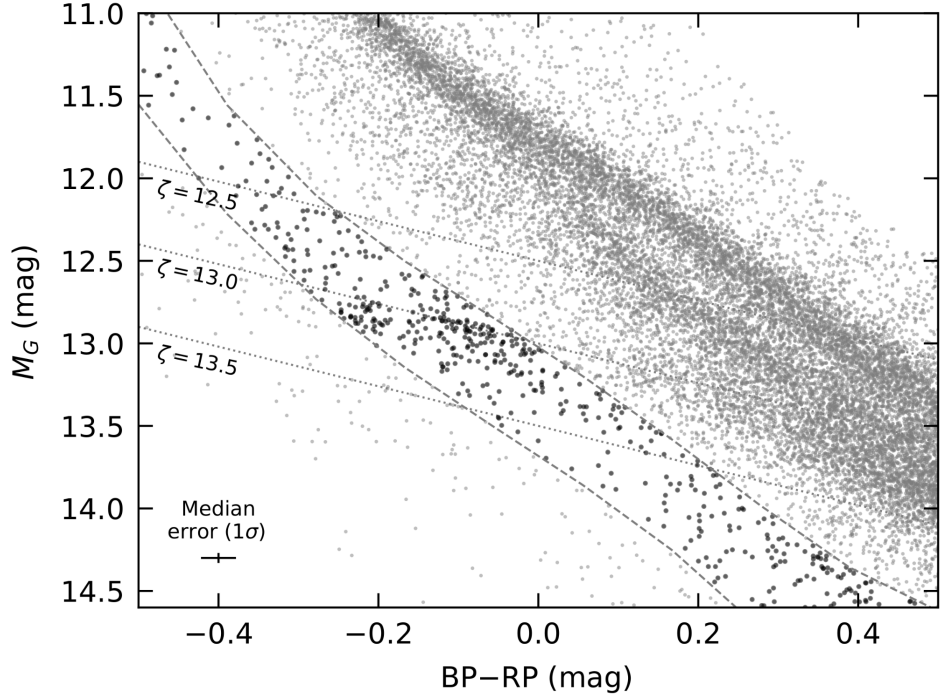

**Fig. 1 | Observational *Gaia* color-magnitude H–R diagram of white dwarfs within 150 pc.** The figure is centered on the *Q*-branch overdensity at $M_G \simeq 13$. Two dashed gray curves delimit the area considered in our analysis, which corresponds to the region occupied by H-atmosphere white dwarfs with masses $1.08 \lesssim M_\star/M_\odot \lesssim 1.23$. For clarity, white dwarfs inside that region are highlighted in black. Three lines of constant $\zeta$ are also shown. The median $1\sigma$ error is displayed in the bottom-left corner.

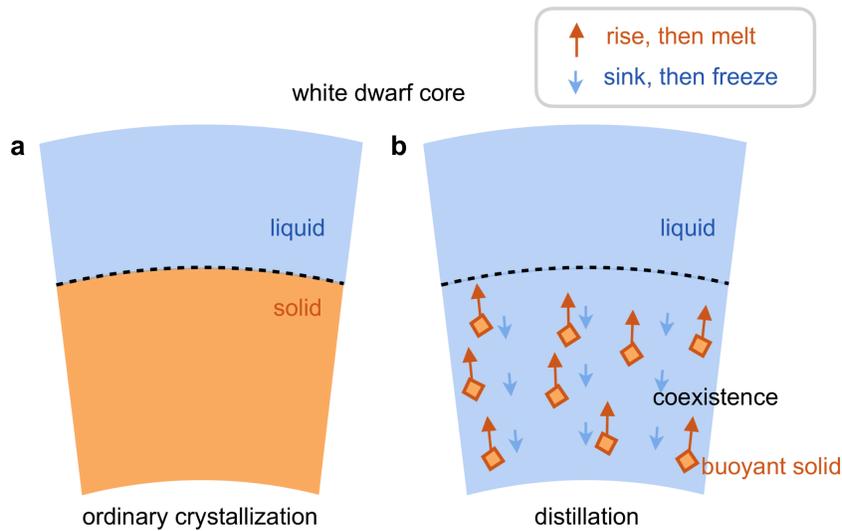

**Fig. 2 | Schematic representation of two scenarios of white dwarf crystallization.** As the star cools down, the solid crystals formed in the canonical case (a) are heavier than the liquid and thus accumulate on the surface of the underlying solid. This crystallization front (dashed line) is the only radial depth that is exactly at the freezing temperature. In contrast, if the solid crystals are lighter than the liquid (b), they float up and eventually melt. As a result, no solid is accumulated, and a large radial range of the white dwarf is in coexistence of floating solid and sinking liquid, forming convection-like flows.

Most high-mass white dwarfs are thought to have O-Ne cores[21], which crystallize following the standard scenario (Fig. 2a)[22]. However, some stellar mergers[10] should produce C-O white dwarfs with considerable amounts of heavy neutron-rich impurities such as $^{22}$Ne, a composition conducive to the formation of buoyant crystals and transforming the canonical crystallization into a distillation process (Fig. 2b)[8]. To show that the large amount of gravitational energy released by distillation can explain all the features of the $Q$ branch, we computed new cooling models[23] of high-mass (1.00–1.25 $M_\odot$) C-O white dwarfs with a reasonably high $^{22}$Ne content (3% by mass) in which distillation is self-consistently included. We then simulated populations of high-mass white dwarfs comprising 5–9% of such objects[5], with the rest consisting of normal O-Ne white dwarfs[21] without distillation (see Methods). Fig. 3 compares the distributions of the observed and synthetic populations in terms of the quantity $\zeta = M_G - 1.2 \times (BP-RP)$[19], which emphasizes the $Q$-branch overdensity at $13.0 \leq \zeta \leq 13.2$. The simulation matches the observed $Q$ branch with high fidelity. In particular, this is the first time that the position, amplitude, and width of the $Q$ branch are properly reproduced[4,18,19]. For comparison, Fig. 3 also shows the distribution obtained from a simulation where distillation is omitted, which completely lacks the sharp peak.

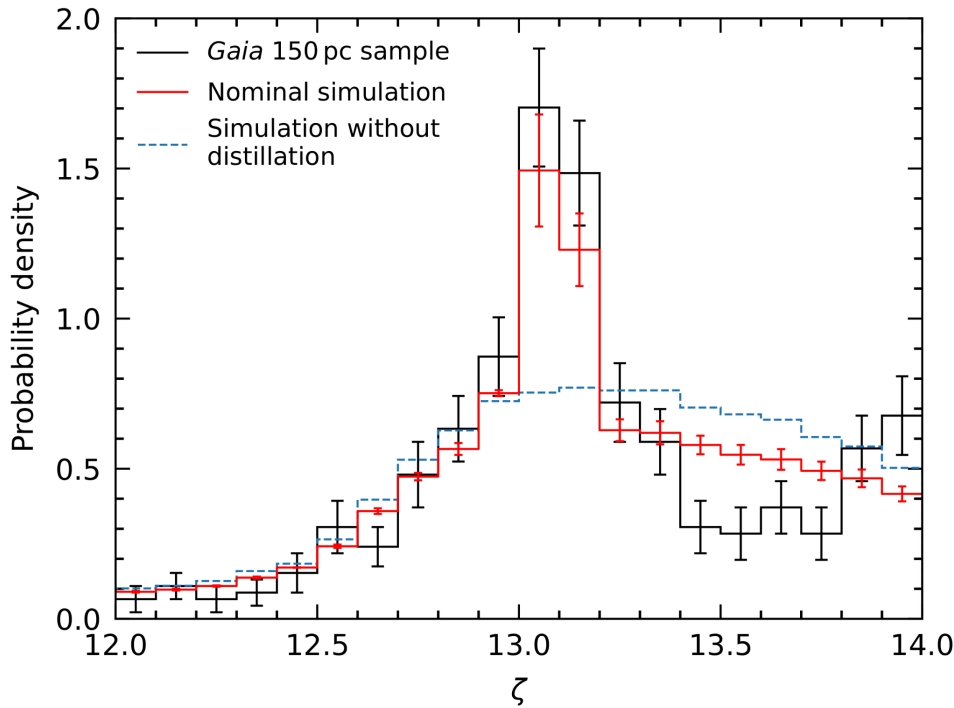

**Fig. 3 | Observed and simulated distributions of high-mass white dwarfs along their cooling tracks.** The histograms show the $\zeta$ distributions of the *Gaia* 150 pc sample (black), of a baseline simulation containing only O-Ne white dwarfs (blue), and of a simulation where 5–9% of stars are assumed to have a C-O core and undergo distillation (red). Only objects located between the two dashed curves of Fig. 1 are considered. The histograms represent probability densities: each bin displays the count of objects within the bin divided by the total number of counts and the bin width. Error bars on the *Gaia* sample histogram show 1$\sigma$ uncertainties based on Poisson statistics in each bin. Error bars on the red histogram display the range of possible outcomes when the fraction of C-O white dwarfs is varied between 5 and 9%. Given this 5–9% range, the simulation with distillation reproduces the observed $Q$-branch peak to within 1$\sigma$. In contrast, the baseline simulation is discrepant at the > 4$\sigma$ level in that same region of $13.0 \leq \zeta \leq 13.2$.

In addition to the agreement with the stellar count in the H–R diagram, the distillation scenario is also consistent with the constraint on the cooling delay time of ≳8 Gyr[5]. Fig. 4a shows the surface luminosity of our distilling C-O white dwarf models as a function of time. The otherwise continuous cooling is almost completely halted for ≃7–13 Gyr (depending on the stellar mass) by the energy release of the distillation process. The higher the mass, the shorter the delay, because a more massive object starts crystallizing and distilling at a higher luminosity and thereby disposes of the extra energy more quickly.

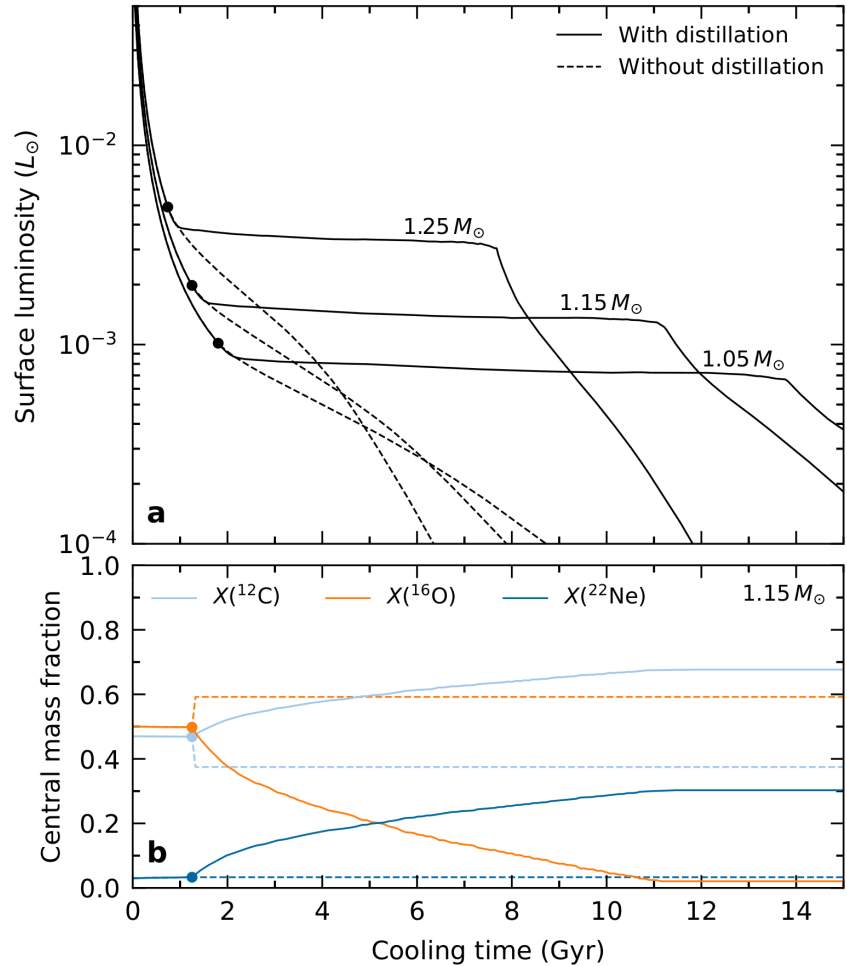

**Fig. 4 | Evolving surface luminosity and central composition of high-mass white dwarf models.** Panel a shows the surface luminosity in 1.05, 1.15, and 1.25 $M_\odot$ models, while panel b shows the central $^{12}$C, $^{16}$O, and $^{22}$Ne mass-fraction abundances in a 1.15 $M_\odot$ model. In both panels, the dashed and solid lines correspond to models ignoring and including $^{22}$Ne distillation, respectively. In all cases, the initial core composition is a uniform mixture with $X(^{12}C) = 0.47$, $X(^{16}O) = 0.50$ and $X(^{22}Ne) = 0.03$, and the envelope has $M(H) = 10^{-10} M_\star$ and $M(He) = 10^{-6} M_\star$. Circles indicate the time at which crystallization and distillation start.

The energy release is the observable manifestation of the extremely efficient chemical transport enabled by distillation. Fig. 4b shows the evolution of the central elemental abundances of a 1.15 $M_\odot$ white dwarf. The distillation mechanism causes the $^{22}$Ne mass fraction to increase from 3% to a final value of ≃30% (determined by the phase diagram[8,24]) in ≃10 Gyr. This high transport efficiency results from the macroscopic flows driven by the buoyant crystals (Fig. 2b), which act much faster than standard microscopic diffusion[18,25,26]. The sedimentation of neutron-rich material in turn leads to a ≃8% increase in central mass density and thus a concomitant release of gravitational energy. The global stellar structure is also affected: the radius of the star decreases by ≃1%, which represents a sizable fraction of the residual cooling-induced contraction of high-mass white dwarfs[27].

The initial $^{22}$Ne mass fraction of 3% assumed in our distilling white dwarf models is consistent with the expected outcome of nuclear burning in some binary merger scenarios[10]. We found that the simulated *Q*-branch overdensity is weakly affected by the initial $^{22}$Ne abundance and remains in agreement with the observations over a wide range of reasonable assumptions (see Methods). Another important characteristic of our white dwarf models is the assumed mass of the residual He layer atop the C-O core, $M(He) = 10^{-6} M_\star$, which is much lower than canonical values for high-mass white dwarfs, $M(He) = 10^{-3}$–$10^{-4} M_\star$[21,28]. This is in line with the observation that part of the *Q*-branch population is severely He deficient[9,29], which has been interpreted as evidence that these stars are merger remnants. We found that assuming a standard He mass yields a poor match to both the location and amplitude of the *Q* branch (see Methods), and thus our results provide independent support for a strong He deficiency. This further solidifies the claim that the delayed population is constituted of merger remnants, as single-star evolution cannot produce the required envelope composition[21,30].

For around ten billion years, the steady energy output of ~$10^{-3}$ $L_\odot$ provided by distillation exceeds the nuclear energy production rate of some main-sequence M dwarfs[31]. This mechanism was first predicted more than thirty years ago[6] but had never been validated by observational evidence. This second stellar life has important implications for age-dating applications, where white dwarfs are used to infer the stellar formation history of the Milky Way[32-35]. A proper account of the effects of distillation is needed to reliably determine the cooling ages of high-mass white dwarfs, which are considered the most direct indicators of past stellar formation given their short main-sequence lifetimes[36]. Distillation has also been predicted to occur in lower-mass white dwarfs with standard compositions[7,8], causing cooling delays of ~1 Gyr that would increase age estimates for most old stellar remnants. Finding conclusive evidence for distillation in a segment of the white dwarf population bolsters the likelihood of this theoretical prediction proving correct.

## References

[1] Althaus, L. G., Córsico, A. H., Isern, J. & García-Berro, E. Evolutionary and pulsational properties of white dwarf stars. *Astron. Astrophys. Rev.* **18** (4), 471–566 (2010).
[2] Saumon, D., Blouin, S. & Tremblay, P.-E. Current challenges in the physics of white dwarf stars. *Phys. Rep.* **988**, 1–63 (2022).
[3] van Horn, H. M. Crystallization of White Dwarfs. *Astrophys. J.* **151**, 227 (1968).
[4] Tremblay, P.-E. *et al.* Core crystallization and pile-up in the cooling sequence of evolving white dwarfs. *Nature* **565** (7738), 202–205 (2019).
[5] Cheng, S., Cummings, J. D. & Ménard, B. A Cooling Anomaly of High-mass White Dwarfs. *Astrophys. J.* **886** (2), 100 (2019).
[6] Isern, J., Hernanz, M., Mochkovitch, R. & García-Berro, E. The role of the minor chemical species in the cooling of white dwarfs. *Astron. Astrophys.* **241** (1), L29–L32 (1991).
[7] Segretain, L. Three-body crystallization diagrams and the cooling of white dwarfs. *Astron. Astrophys.* **310**, 485–488 (1996).
[8] Blouin, S., Daligault, J. & Saumon, D. $^{22}$Ne Phase Separation as a Solution to the Ultramassive White Dwarf Cooling Anomaly. *Astrophys. J. Lett.* **911** (1), L5 (2021).
[9] Hollands, M. A. *et al.* An ultra-massive white dwarf with a mixed hydrogen-carbon atmosphere as a likely merger remnant. *Nature Astronomy* **4**, 663–669 (2020).

10. Shen, K. J., Blouin, S. & Breivik, K. The Q Branch Cooling Anomaly Can Be Explained by Mergers of White Dwarfs and Subgiant Stars. *Astrophys. J. Lett.* **955** (2), L33 (2023).

11. Winget, D. E. *et al.* An Independent Method for Determining the Age of the Universe. *Astrophys. J. Lett.* **315**, L77 (1987).

12. Fontaine, G., Brassard, P. & Bergeron, P. The Potential of White Dwarf Cosmochronology. *Publ. Astron. Soc. Pac.* **113** (782), 409–435 (2001).

13. Gaia Collaboration. Gaia Data Release 2. Observational Hertzsprung-Russell diagrams. *Astron. Astrophys.* **616**, A10 (2018).

14. Althaus, L. G., García-Berro, E., Isern, J., Córsico, A. H. & Miller Bertolami, M. M. New phase diagrams for dense carbon-oxygen mixtures and white dwarf evolution. *Astron. Astrophys.* **537**, A33 (2012).

15. Blouin, S., Daligault, J., Saumon, D., Bédard, A. & Brassard, P. Toward precision cosmochronology. A new C/O phase diagram for white dwarfs. *Astron. Astrophys.* **640**, L11 (2020).

16. Bauer, E. B. Carbon-Oxygen Phase Separation in Modules for Experiments in Stellar Astrophysics (MESA) White Dwarf Models. *Astrophys. J.* **950** (2), 115 (2023).

17. Kilic, M. *et al.* The 100 pc White Dwarf Sample in the SDSS Footprint. *Astrophys. J.* **898** (1), 84 (2020).

18. Bauer, E. B., Schwab, J., Bildsten, L. & Cheng, S. Multi-gigayear White Dwarf Cooling Delays from Clustering-enhanced Gravitational Sedimentation. *Astrophys. J.* **902** (2), 93 (2020).

19. Camisassa, M. E. *et al.* Forever young white dwarfs: When stellar ageing stops. *Astron. Astrophys.* **649**, L7 (2021).

20. Fleury, L., Caiazzo, I. & Heyl, J. The cooling of massive white dwarfs from Gaia EDR3. *Mon. Not. R. Astron. Soc.* **511** (4), 5984–5993 (2022).

21. Camisassa, M. E. et al. The evolution of ultra-massive white dwarfs. *Astron. Astrophys.* **625**, A87 (2019).

22. Blouin, S. & Daligault, J. Phase Separation in Ultramassive White Dwarfs. *Astrophys. J.* **919** (2), 87 (2021).

23. Bédard, A., Brassard, P., Bergeron, P. & Blouin, S. On the Spectral Evolution of Hot White Dwarf Stars. II. Time-dependent Simulations of Element Transport in Evolving White Dwarfs with STELUM. *Astrophys. J.* **927** (1), 128 (2022).

24. Caplan, M. E., Horowitz, C. J. & Cumming, A. Neon Cluster Formation and Phase Separation during White Dwarf Cooling. *Astrophys. J. Lett.* **902** (2), L44 (2020).

25. Deloye, C. J. & Bildsten, L. Gravitational Settling of $^{22}$Ne in Liquid White Dwarf Interiors: Cooling and Seismological Effects. *Astrophys. J.* **580** (2), 1077–1090 (2002).

26. García-Berro, E., Althaus, L. G., Córsico, A. H. & Isern, J. Gravitational Settling of $^{22}$Ne and White Dwarf Evolution. Astrophys. J. 677 (1), 473–482 (2008).

27. Bédard, A., Bergeron, P., Brassard, P. & Fontaine, G. On the Spectral Evolution of Hot White Dwarf Stars. I. A Detailed Model Atmosphere Analysis of Hot White Dwarfs from SDSS DR12. *Astrophys. J.* **901** (2), 93 (2020).

28. Romero, A. D., Kepler, S. O., Córsico, A. H., Althaus, L. G. & Fraga,L. Asteroseismological Study of Massive ZZ Ceti Stars with Fully Evolutionary Models. *Astrophys. J.* **779** (1), 58 (2013).

29. Koester, D., Kepler, S. O. & Irwin, A. W. New white dwarf envelope models and diffusion. Application to DQ white dwarfs. *Astron. Astrophys.* **635**, A103 (2020).

30. Althaus, L. G. *et al.* The formation of ultra-massive carbon-oxygen core white dwarfs and their evolutionary and pulsational properties. *Astron. Astrophys.* **646**, A30 (2021).

31. Zombeck, M. V. *Handbook of Space Astronomy and Astrophysics* 3rd edn (Cambridge University Press, Cambridge, 2006).

32. García-Berro, E. *et al.* A white dwarf cooling age of 8 Gyr for NGC 6791 from physical separation processes. *Nature* **465** (7295), 194–196 (2010).

33. Kilic, M. *et al.* The Ages of the Thin Disk, Thick Disk, and the Halo from Nearby White Dwarfs. *Astrophys. J.* **837** (2), 162 (2017).

34. Fantin, N. J. *et al.* The Canada-France Imaging Survey: Reconstructing the Milky Way Star Formation History from Its White Dwarf Population. *Astrophys. J.* **887** (2), 148 (2019).

35. Cukanovaite, E. *et al.* Local stellar formation history from the 40 pc white dwarf sample. *Mon. Not. R. Astron. Soc.* **522** (2), 1643–1661 (2023).

36. Isern, J. The Star Formation History in the Solar Neighborhood as Told by Massive White Dwarfs. *Astrophys. J. Lett.* **878** (1), L11 (2019).

**Methods**

**Cooling models: basic assumptions.** Our C-O white dwarf cooling models were computed using the STELUM code[23]. We generated and evolved 1.00–1.25 $M_\odot$ model structures consisting of a $^{12}$C-$^{16}$O-$^{22}$Ne core surrounded by a He mantle and an outermost H layer. We adopted envelope layer masses M(H) = $10^{-10}$ $M_\star$ and M(He) = $10^{-6}$ $M_\star$ along with an initially uniform core with mass-fraction abundances $X(^{16}O)$ = 0.50 and $X(^{22}Ne)$ = 0.03. We henceforth refer to this as our post-merger chemical structure. The envelope composition is based on empirical characterization of $Q$-branch white dwarfs[9,29] and is also consistent with the expected outcome of nuclear burning following a stellar merger[30]. The core composition is taken from recent theoretical work on white dwarf-subgiant mergers[10], with the exception that $^{22}$Ne is used as a proxy for all neutron-rich species. This choice is motivated by the fact that white dwarf-subgiant mergers represent the most promising scenario for the formation of ultramassive white dwarfs with both a C-O core and a high content of neutron-rich impurities. The uniform composition is a direct consequence of the merger event, which generates thermohaline mixing that homogenizes the entire core[10]. This also explains the slightly lower central $^{16}$O abundance compared to standard values expected from single-star evolution[16,37], $X(^{16}O) \simeq 0.55$–0.60, as the $^{16}$O normally concentrated towards the center is homogeneously redistributed over the core.

We ignored general relativistic effects, which is an excellent approximation within the stellar mass range considered[38,39]. We also omitted rotation and magnetic fields; although some merger remnants are observed to be rapidly rotating and/or strongly magnetic[40,41], the effects on the stellar structure and evolution are expected to be negligible[2,42]. We employed the radiative opacities of the OPAL project[43] and the latest conductive opacities[44,45]. Convective energy transport in the envelope was handled following the ML2 version of the mixing-length theory[46,47]. We used a simple gray atmosphere as outer boundary condition, a valid approximation for $Q$-branch white dwarfs as they have not yet undergone surface-core convective coupling[12,48].

Element transport was modeled through the usual scheme in which the time-dependent transport equations are fully coupled to the stellar structure equations[23]. Standard microscopic diffusion was included in the gas and liquid phases using newly implemented diffusion coefficients[49,50]. This allows for a state-of-the-art description of the heating effect due to $^{22}$Ne diffusion in the core[18,26,51]. We considered gravitational settling and chemical diffusion but ignored thermal diffusion, as it is negligible in the nearly isothermal core of white dwarfs. We omitted chemical mixing due to convection in the envelope, because this process can give rise to severe numerical difficulties[52]. This assumption is certainly valid for the DA-type members of the $Q$-branch population, whose H-dominated atmosphere indicates that the superficial H layer has not been altered by convection. This is however not true for their DQ-type counterparts, which likely owe their H-deficient surface composition to the convective mixing of the outer H, He, and C layers[29,52]. Nevertheless, we verified that artificially imposing a uniform He-C envelope with typical abundances[29,53] barely affects the evolution, whether before, during, or after the distillation phase. The overall opacity turns out to be similar to that provided by a thin pure-H layer, which thus represents an acceptable approximation even for the DQ stars.

**Cooling models: crystallization and distillation.** As a prerequisite to the inclusion of the distillation mechanism in STELUM, we first upgraded the treatment of standard C-O fractionation upon crystallization in the absence of neutron-rich impurities. We replaced the approximate heating term used previously[23,54] with a detailed description of the composition changes in the liquid and solid regions, similar to that adopted in the MESA code[16]. This procedure involves the following steps, which are performed at the end of each evolutionary time step.

1. Identifying the layers that have solidified during the last time step. The phase transition is assumed to occur when the plasma coupling parameter $\Gamma$ becomes larger than the critical value $\Gamma_{cr}$ predicted by the C-O phase diagram of ref. 55. Note that the coupling parameter of the plasma mixture is defined as $\Gamma = \langle Z^{5/3} \rangle e^2 / a_e k_B T$, where $\langle Z^{5/3} \rangle = \sum_i Z_i^{5/3} x_i$ ($x_i$ being the number fraction of the ionic species of charge $Z_i$), $e$ is the elementary charge, $a_e$ is the radius of a sphere whose volume corresponds to the mean volume per electron, $k_B$ is the Boltzmann constant, and $T$ is the temperature.
2. Adjusting the composition of the newly crystallized layers. The $^{16}O$ abundance is increased following the same phase diagram as above, while the $^{12}C$ abundance is decreased by an equal amount. The composition of the solid region is then held fixed for the rest of the evolutionary calculation.
3. Adjusting the composition of the innermost liquid layer to enforce total elemental mass conservation. Given the abundance changes in the solid region, the first liquid layer above the crystallization front is enriched in $^{12}C$ and depleted in $^{16}O$, which produces an inverted molecular weight gradient.
4. Imposing a uniform composition in the dynamically unstable liquid layers. The instability arising from the molecular weight gradient is assumed to mix material outward in a homogeneous and instantaneous way[16,56-58]. The extent of the mixed region is found by redistributing the elements over an increasing number of layers until the Ledoux stability criterion is satisfied[16,58,59].

In the case of $^{22}Ne$ distillation, fractionation produces $^{22}Ne$-poor crystals that float up and eventually melt, thereby displacing $^{22}Ne$-rich liquid downward. The net result is that the central crystal-forming region remains globally liquid (hence the term crystal-forming rather than crystallized) and gradually becomes enriched in $^{12}C$ and $^{22}Ne$ and depleted in $^{16}O$ at the expense of the outer layers[8]. We implemented this process in STELUM using the same algorithm as above but with a few modifications to steps 1 and 2. In step 1, the impact of $^{22}Ne$ on the onset of crystallization is taken into account through a new analytical fit to the high-resolution C-O-Ne phase diagram of ref. 8. The fitting formula specifies the critical coupling parameter of C, $\Gamma_{cr,C}$, and is expressed as a correction to the C-O phase diagram of ref. 55, $\Gamma_{cr,C} = \Gamma_{cr,C}^0 + (c_1 x_O + c_2 x_O^2 + c_3 x_O^3) x_{Ne}$, where $\Gamma_{cr,C}^0$ is the uncorrected value, $x_O$ and $x_{Ne}$ are the number fractions of $^{16}O$ and $^{22}Ne$, and the numerical coefficients are $c_1 = 1096.69$, $c_2 = -3410.33$, and $c_3 = 2408.44$. Because the C-O-Ne phase diagram of ref. 8 is limited to relatively small $^{22}Ne$ abundances, the above expression is only valid for $x_{Ne} < 0.04$. Furthermore, the crystal-forming layers so identified are assumed to remain liquid, meaning that their composition is allowed to change at each subsequent time step. In step 2, the chemical evolution of the crystal-forming region is handled using a prescription for the elemental abundances as a function of the coupling parameter of C, $\Gamma_C$. Current phase diagrams specify the composition at the beginning and end of distillation[8,24], but the exact trajectory followed between these two points remains unknown. We opted for the simplest possible prescription, where the abundances change linearly with $\Gamma_C$ between the initial and final states. The latter is given by $(\Gamma_C, x_C, x_O, x_{Ne}) = (208, 0.8, 0.0, 0.2)$ in terms of number fractions, corresponding to a $^{22}Ne$ mass fraction $X(^{22}Ne) \simeq 0.31^8$. Once the composition of the crystal-forming layers has been updated, steps 3 and 4 are applied as described above, except that $^{22}Ne$ mass conservation and redistribution are also considered. The energy released by

$^{22}$Ne distillation is naturally taken into account through the coupling of the chemical transport and stellar structure equations in the STELUM code[23].

The distillation process is assumed to stop once the whole $^{22}$Ne content of the star has been transported to the central crystal-forming region. This is admittedly an uncertain aspect of our modeling, as it is entirely possible that distillation terminates earlier; this point is addressed in "Sensitivity to the distillation implementation" below. At this stage, the outer $^{22}$Ne-free layers of the core are expected to start crystallizing normally, as there is no more $^{22}$Ne to produce a density reversal with respect to the overlying liquid. These stable solid layers presumably block the floating motion of the buoyant crystals immediately underneath, which thus remain in place and in turn obstruct crystals originating from deeper layers, and so on. The likely outcome is that the whole $^{22}$Ne-rich region solidifies at once, thereby completely halting the distillation process. For this reason, in our models, the inner $^{22}$Ne-rich core is assumed to freeze in place at the end of distillation, and its composition is held fixed afterwards. This region typically encompasses ≃25% of the total mass of the star; the sudden solidification of that much mass releases a large amount of latent heat, which can cause significant numerical instabilities. To avoid such instabilities, the progression of the crystallization front is artificially slowed down by setting the critical coupling parameter of the entire core to its value at the center (where it is highest as the $^{22}$Ne abundance is highest[24,60]). This ensures that the energy release is more gradual in both space and time, thereby greatly improving numerical convergence. This procedure slightly accelerates the post-distillation cooling of our models, but this is of no consequence to our analysis given that most delayed white dwarfs are not old enough to have reached this phase (Fig. 4a). Once the crystallization front reaches the outer $^{22}$Ne-free layers, standard C-O fractionation occurs as described above.

To better illustrate our modeling assumptions, we show in Extended Data Figure 1 the composition profile of our 1.15 M$_\odot$ model just before, during, and just after distillation, as well as at the very end of the evolutionary sequence (at which point nearly the entire core has solidified). The distillation mechanism produces an increasingly $^{22}$Ne-rich, $^{16}$O-poor inner core, which also expands with time as more layers start forming buoyant crystals. In this region, the abundance profiles simply reflect the $\Gamma_C$ profile of the star[2] as a result of our chemical evolution prescription. At the end of distillation, there is no more $^{22}$Ne in the outer portion of the core, and the composition profile of the inner $^{22}$Ne-rich layers ($m/M_\star \lesssim 0.25$) is subsequently held fixed. In the $^{22}$Ne-free region, the final $^{16}$O abundance profile is the well-known result of standard fractionation upon crystallization[14,16].

During the distillation process, the gravitational energy released almost exactly balances the energy radiated away, such that the surface luminosity and temperature remain essentially constant. In practice, the use of a finite time step in our numerical calculations gives rise to small oscillations around the mean values (typically ~1% in $T_{eff}$). We mitigated these numerical artifacts by applying a simple moving-average smoothing procedure (with an averaging window of ~1 Gyr) along the luminosity/temperature plateau of our evolutionary sequences. This is important to ensure an accurate and well-behaved interpolation of the cooling models in the population synthesis simulations. All cooling sequences used and shown in this paper are the final smoothed versions.

A key characteristic of the observed $Q$ branch reproduced by our calculations is the fact that it is much narrower than the luminosity range of crystallization itself[5]. This is explained by two properties of the distillation mechanism. First, distillation operates only in the initial phase of the crystallization process, as it must stop once all neutron-rich impurities have been displaced to the central layers. Second, the rate of energy release during the distillation process itself is not uniform. It is roughly proportional to the total mass of the coexistence region, and therefore peaks only after distillation has operated for some time. This explains the offset between the start of distillation and the cooling pause in Fig. 4a.

**Population synthesis.** We considered three distinct ultramassive white dwarf populations in our population synthesis: 5–9% of C-O white dwarfs that undergo distillation, ≃30% of O-Ne white dwarfs with merger delays, and the remaining standard O-Ne white dwarfs. The first group is characterized by a composition similar to that of white dwarf-subgiant merger remnants[10], although we do not exclude other evolutionary channels. We assigned a proportion of 5–9% to this population as constrained by a previous kinematic analysis[5]. These objects were modeled using the cooling calculations detailed above. The second group consists of double white dwarf merger products. We supposed that it accounts for ≃30% of the ultramassive population (with a small dependence on the mass)[61], and that they harbor O-Ne cores[62]. Finally, the remainder are standard O-Ne white dwarfs resulting from single-star evolution and were modeled using existing cooling tracks[21].

We assumed that the age distribution of stars in the 150 pc solar neighborhood is constant between 0 and 10.5 Gyr[35]. We did not explicitly model the initial mass function, because a sizable fraction of the ultramassive population emerged from stellar mergers. Instead, we used a large sample of white dwarfs[17] to infer a $\propto M_\star^{-2}$ empirical probability distribution function for white dwarfs more massive than 1 $M_\odot$. For a given white dwarf mass taken from this distribution, we then added the pre-white dwarf lifetime[63,64] and a merger delay when applicable. We modeled the delay-time distribution using a log-normal distribution. For the natural logarithm of ages expressed in Gyr, $\mu = 0.5$ and $\sigma = 0.7$, which corresponds to an average delay of 2.1 Gyr[61]. The *Gaia* magnitudes were then calculated using pure-H atmosphere models[65,66], except for 50% of the extra-delayed C-O core population, for which we instead used He-dominated, C-polluted atmosphere models with $N(C)/N(He) = 0.1$[53]. This assumption is motivated by a kinematic analysis of spectroscopically observed *Q*-branch white dwarfs, which revealed that the extra-delayed population is composed of DA and DQ stars in similar proportions while the remainder comprises very few (if any) DQ stars[5]. For consistency with previous work[5], we only considered ultramassive white dwarfs whose *Gaia* photometry indicates a mass comprised in the 1.08–1.23 $M_\odot$ interval when modeled assuming H atmospheres and O-Ne cores.

**Model atmospheres.** An existing DQ model atmosphere grid[53] was extended to calculate the *Gaia* magnitudes of *Q*-branch objects with C-polluted photospheres. The original model grid stopped at $T_{\rm eff} = 16{,}000$ K and $\log g = 9.0$; we expanded it to $T_{\rm eff} = 20{,}000$ K and $\log g = 9.5$ using the same model atmosphere code[67,68].

**Sample selection.** The white dwarf sample shown in Figs. 1 and 3 was taken from the *Gaia* EDR3 white dwarf catalog[69]. We selected all high-confidence white dwarf candidates (i.e., those with a probability of being white dwarfs of at least 90%, $P_{\rm WD} \geq 0.9$) that lie within $D = 150$ pc of the Sun. This distance cut-off is justified in Extended Data Figure 2, where we show the cumulative distribution functions (CDFs) of the distances of high-confidence white dwarf candidates for three different $M_G$ bins. For the faintest bin relevant to our analysis ($13.5 < M_G \leq 14.5$), the CDF departs from the expected $\propto D^3$ relation at ≃150 pc.

This indicates that the sample starts to be noticeably incomplete at faint magnitudes past 150 pc. Accordingly, we only selected stars within 150 pc to avoid biasing the sample in favor of brighter stars.

We cross-matched all 458 white dwarfs from the 150 pc sample that are used in our analysis (i.e., those falling between the gray dashed lines in Fig. 1) with the Montreal White Dwarf Database[70]. We found that only 3 out of the 79 objects with a known spectral type are of the DB class, with DAs and DQs making up most of the remainder. This justifies our use of pure-H atmosphere models for all objects in our simulations, except for half of the extra-delayed C-O core population[5], which we modeled using DQ model atmospheres.

**Sensitivity to the assumed composition.** As mentioned above, our nominal cooling sequences assume a post-merger chemical structure, with an extremely thin envelope and an initially uniform core. To investigate the impact of these assumptions on our results, we also computed a set of white dwarf models with more standard envelope and core stratifications (that is, emulating the predictions of single-star evolution calculations). For this test, we used $M(H) = 10^{-6}\,M_\star$, $M(He) = 10^{-3}\,M_\star$[21,28], and the nonuniform C-O profile of the 1.0 $M_\odot$ model of ref. 16 (but still assuming $X(^{22}Ne) = 0.03$). The resulting cooling sequences (Extended Data Figure 3a) are significantly different from our nominal case: for a given mass, the constant-luminosity phase due to $^{22}$Ne distillation occurs at a higher luminosity and is thus shorter. The reason is that the models start crystallizing earlier, a behavior due mainly to the larger He layer mass, but also to the higher central O abundance. In particular, the larger amount of He increases the overall transparency of the envelope, resulting in a higher surface luminosity for a given core temperature. The earlier and shorter cooling delays lead to a much poorer agreement between the simulated and observed $Q$-branch overdensities, both in terms of location and amplitude (Extended Data Fig. 4a). The thin He layer of our post-merger-type models also explains the longer distillation cooling delays compared to previous estimates[8].

Another potentially important parameter for our study is the initial $^{22}$Ne abundance, which dictates the amount of gravitational energy available for release through distillation. Given current uncertainties on this quantity[10], we computed additional cooling sequences assuming a lower $^{22}$Ne content of $X(^{22}Ne) = 0.025$ along with our default post-merger chemical structure. For a given mass, the time evolution of the surface luminosity is qualitatively similar to the nominal case, but the cooling delay due to distillation is unsurprisingly smaller (≈6–11 Gyr for 1.00–1.25 $M_\odot$). Repeating our population synthesis simulation with these models, we find that the results are only weakly affected (Extended Data Figure 4b). Similarly, we found that increasing the $^{22}$Ne content to $X(^{22}Ne) = 0.04$ has a negligible impact on the predicted $Q$ branch. The cooling delays for $X(^{22}Ne) = 0.03$ are already long enough that the vast majority of delayed white dwarfs are still stuck on the $Q$ branch at the final 10.5 Gyr age.

**Sensitivity to the distillation implementation.** The C-O-Ne phase diagram predicts that the solid phase is buoyant as long as the abundance of $^{22}$Ne in the liquid remains above a well-defined threshold[8]. As the distillation process unfolds, there is inevitably a point where the $^{22}$Ne abundance in the outermost crystal-forming layer reaches this critical value. What happens past that juncture remains unclear. A priori, distillation could stop because the solid phase is no longer buoyant. However, despite not being buoyant, the solid phase is still depleted in $^{22}$Ne. This means that crystallization should increase the $^{22}$Ne abundance in the liquid, thereby reinstating the conditions required for distillation. Stopping distillation would then be an unstable state. To address this uncertainty, we calculated additional cooling sequences corresponding to an extreme scenario where distillation is terminated when the threshold abundance is first reached (Extended Data Figure 3b). The distillation-induced cooling delay is then shorter, but it remains sufficient to explain the $Q$-branch overdensity (Extended Data Figure 5). Our conclusions are therefore insensitive to the treatment of distillation past this critical juncture.

**Microscopic diffusion of $^{22}$Ne.** It has been claimed that simple microscopic diffusion (often referred to as gravitational settling) of $^{22}$Ne in C-O white dwarfs can possibly explain the $Q$-branch cooling anomaly[19], an idea similar to the "sedimentars'" proposed two decades earlier[71]. However, this mechanism cannot reproduce the narrowness of the $Q$ branch (see Fig. 4 of ref. 19). Moreover, to obtain a long enough delay time, a very large $^{22}$Ne mass fraction of 0.06 is required, a value that far surpasses what can be explained by any known evolutionary pathway. This scenario also requires that as many as ~50% of all ultramassive white dwarfs experience a delayed cooling, in sharp disagreement with kinematic constraints[5]. In short, this is not a viable scenario.

Our cooling models are consistent with these conclusions. Ignoring distillation and assuming $X(^{22}\text{Ne}) = 0.06$ (along with our usual post-merger composition), we find that the slowdown of the cooling process is much less pronounced than in our default setup (Extended Data Figure 3c). Gravitational settling is simply far less efficient than distillation at transporting neutron-rich species to the center of the star (Extended Data Figure 6). We note that the settling-induced cooling delays of our $X(^{22}\text{Ne}) = 0.06$ models are somewhat shorter than those reported by ref. 19, in line with independent calculations[16]. A population synthesis simulation where 5–9% of ultramassive white dwarfs follow this evolutionary scenario yields a small bump centered on the $Q$ branch, but this overdensity is too shallow and wide to reproduce the observations (Extended Data Figure 7). This is a direct consequence of the fact that, unlike distillation, gravitational settling operates over a wide range of luminosities (Extended Data Figure 3c).

**Sensitivity to star formation history.** We assumed a uniform stellar age distribution for our 150 pc solar neighborhood sample for up to 10.5 Gyr. This distribution is both consistent with the recent analysis of a nearby white dwarf sample[35] and expected due to the counterbalancing of the declining star formation rate[72] by the thinner disk scale-height for younger stars[73]. However, independent studies instead find strong variations in the effective star formation history of the Galactic disk, with a stellar formation burst 2–3 Gyr ago[74]. To verify whether this could have an effect on our conclusions[20], we show in Extended Data Figure 8 the results of population synthesis simulations identical to those presented in Fig. 3 but using a non-uniform age distribution inspired by the results of ref. 74. The age distribution is assumed to correspond to the sum of a uniform distribution and a Gaussian function centered at 2.5 Gyr ago with $\sigma = 1$ Gyr and an amplitude that increases the number of 2.5 Gyr-old stars by 60% compared to the uniform baseline. A small bump associated with the stellar formation burst appears just after the $Q$-branch peak, but its amplitude and location do not match the $Q$-branch signal. The agreement between our nominal simulation and the *Gaia* data has worsened compared to Fig. 3, but it is clear that distillation is still needed to explain the sharp peak between $\zeta = 13.0$ and $13.2$. Our conclusions are therefore insensitive to the assumed stellar formation history.

**Sensitivity to merger delay.** The log-normal distribution for merger time delays used in our population synthesis mimics the results of ref. 61 for double white dwarf mergers. However, this distribution may not be appropriate for all types of mergers that can contribute to the ultramassive population. We repeated the population synthesis simulations of Fig. 3 assuming no merger delay for all stars and found that our results are barely affected (Extended Data Figure 9a). Another test was performed where the delay times were doubled for all merger remnants, and the same conclusion was reached (Extended Data Figure 9b). Our results are therefore not sensitive to the assumed merger delay time distribution. This is expected because a merger creates only an overall shift of cooling times but no pile-up along the cooling track.

**Population density at $\zeta > 13.4$.** A tension between the observed and simulated populations appears at $\zeta > 13.4$ in Fig. 3. In the $13.4 < \zeta < 13.8$ range, this may be due to the high conductive opacities assumed in the O-Ne white dwarf models[21] used for the bulk of the synthetic population[45,75]. At $\zeta > 13.8$, the observed population is likely contaminated by faint lower-mass white dwarfs with uncertain *Gaia* data (see the bottom-right corner of Fig. 1). In any case, this tension does not impact our conclusions regarding the *Q* branch and distillation, which are based on the analysis of white dwarfs with $\zeta < 13.4$.

**Kinematics of *Q*-branch white dwarfs.** It has been claimed that the velocity dispersion of white dwarfs on the *Q* branch in the direction of Galactic rotation is higher than even the oldest population of the Galactic disk[76]. However, we believe that is only an apparent mismatch due to the invalidity of fitting a Gaussian tail to a non-Gaussian velocity distribution when its dispersion is large. Below we show that the *Q*-branch kinematics matches that of stars in the solar vicinity with ages older than 1–2 Gyr, which is a prediction from the cooling delay scenario.

To avoid inconsistencies caused by selection effects in different datasets used in the literature, we directly compared the velocity distribution of *Q*-branch white dwarfs to a sample of M dwarfs in *Gaia*, with the same distance cut of 150 pc from the Sun and a color cut of $1.49 \leq \text{BP-RP} \leq 1.50$. The absolute magnitude of this sample is higher than the *Q* branch ($M_G \simeq 7.5$), and thus it is also complete at the same distance. In a cooling delay scenario, the *Q* branch is expected to have the same age distribution as M dwarfs older than 1–2 Gyr, because both samples have no age bias or selection in that range. We computed the velocity dispersion of both samples projected in the Galactic radial, rotation, and vertical directions, after correcting for the contribution from young objects. We first removed low-velocity ($< 20$ km/s) objects, i.e., young stars from both samples. The M dwarf sample then has dispersions of 34, 31, 25 km/s, respectively, while the *Q* branch ($\zeta$ within 13.0–13.2, mass within 1.08–1.23 $M_\odot$) has 31, 30, 23 km/s. The *Q* branch dispersion is somewhat smaller than the M-dwarf sample, but, because the *Q*-branch sample contains a high fraction of normal white dwarfs (about half, as estimated in ref. 5 or read from Fig. 3) that are 1–2 Gyr old, the tail of their velocity distribution may still contaminate the range above 20 km/s. We estimated their contribution using the velocities of white dwarfs with $12.4 < \zeta < 12.8$. After such a correction, the velocity dispersion of the *Q* branch becomes 32, 32, 25 km/s, accurately matching the M dwarf sample and supporting the idea that the *Q* branch is an accumulation of stars with a wide range of ages caused by a long cooling delay. It may be worth further investigating the slightly higher ratio between dispersions along the Galactic rotation and radial directions, but we find no evidence that the *Q*-branch velocity dispersion exceeds the expected disk values.

## Methods references

[37] Salaris, M., Cassisi, S., Pietrinferni, A. & Hidalgo, S. The updated BASTI stellar evolution models and isochrones - III. White dwarfs. *Mon. Not. R. Astron. Soc.* **509** (4), 5197–5208 (2022).

[38] Althaus, L. G. *et al.* Structure and evolution of ultra-massive white dwarfs in general relativity. *Astron. Astrophys.* **668**, A58 (2022).

[39] Althaus, L. G. *et al.* Carbon-oxygen ultra-massive white dwarfs in general relativity. *Mon. Not. R. Astron. Soc.* **523** (3), 4492–4503 (2023).

[40] Williams, K. A., Montgomery, M. H., Winget, D. E., Falcon, R. E. & Bierwagen, M. Variability in Hot Carbon-dominated Atmosphere (Hot DQ) White Dwarfs: Rapid Rotation? *Astrophys. J.* **817** (1), 27 (2016).

[41] Kilic, M. et al. The merger fraction of ultramassive white dwarfs. *Mon. Not. R. Astron. Soc.* **518** (2), 2341–2353 (2023).

[42] Kippenhahn, R., Weigert, A. & Weiss, A. *Stellar Structure and Evolution* 2nd edn (Springer, 2012).

[43] Iglesias, C. A. & Rogers, F. J. Updated Opal Opacities. *Astrophys. J.* **464**, 943 (1996).

[44] Cassisi, S., Potekhin, A. Y., Pietrinferni, A., Catelan, M. & Salaris, M. Updated Electron-Conduction Opacities: The Impact on Low-Mass Stellar Models. *Astrophys. J.* **661** (2), 1094–1104 (2007).

[45] Blouin, S., Shaffer, N. R., Saumon, D. & Starrett, C. E. New Conductive Opacities for White Dwarf Envelopes. *Astrophys. J.* **899** (1), 46 (2020).

[46] Böhm-Vitense, E. Über die Wasserstoffkonvektionszone in Sternen verschiedener Effektivtemperaturen und Leuchtkräfte. Mit 5 Textabbildungen. *Zeitschrift fuer Astrophysik* **46**, 108 (1958).

[47] Tassoul, M., Fontaine, G. & Winget, D. E. Evolutionary Models for Pulsation Studies of White Dwarfs. *Astrophys. J. Suppl. Ser.* **72**, 335 (1990).

[48] Rohrmann, R. D., Althaus, L. G., García-Berro, E., Córsico, A. H. & Miller Bertolami, M. M. Outer boundary conditions for evolving cool white dwarfs. *Astron. Astrophys.* **546**, A119 (2012).

[49] Stanton, L. G. & Murillo, M. S. Ionic transport in high-energy-density matter. *Phys. Rev. E* **93** (4), 043203 (2016).

[50] Caplan, M. E., Bauer, E. B. & Freeman, I. F. Accurate diffusion coefficients for dense white dwarf plasma mixtures. *Mon. Not. R. Astron. Soc.* **513** (1), L52–L56 (2022).

[51] Camisassa, M. E. *et al.* The Effect of $^{22}$Ne Diffusion in the Evolution and Pulsational Properties of White Dwarfs with Solar Metallicity Progenitors. *Astrophys. J.* **823** (2), 158 (2016).

[52] Althaus, L. G., García-Berro, E., Córsico, A. H., Miller Bertolami, M. M. & Romero, A. D. On the Formation of Hot DQ White Dwarfs. *Astrophys. J. Lett.* **693** (1), L23–L26 (2009).

[53] Coutu, S. *et al.* Analysis of Helium-rich White Dwarfs Polluted by Heavy Elements in the Gaia Era. *Astrophys. J.* **885** (1), 74 (2019).

[54] Isern, J., García-Berro, E., Hernanz, M. & Chabrier, G. The Energetics of Crystallizing White Dwarfs Revisited Again. *Astrophys. J.* **528** (1), 397–400 (2000).

[55] Blouin, S. & Daligault, J. Direct evaluation of the phase diagrams of dense multicomponent plasmas by integration of the Clapeyron equations. *Phys. Rev. E* **103** (4), 043204 (2021).

[56] Salaris, M. *et al.* The Cooling of CO White Dwarfs: Influence of the Internal Chemical Distribution. *Astrophys. J.* **486** (1), 413–419 (1997).

[57] Fuentes, J. R., Cumming, A., Castro-Tapia, M. & Anders, E. H. Heat Transport and Convective Velocities in Compositionally Driven Convection in Neutron Star and White Dwarf Interiors. *Astrophys. J.* **950** (1), 73 (2023).

[58] Montgomery, M. H. & Dunlap, B. H. Fluid Mixing during Phase Separation in Crystallizing White Dwarfs. *arXiv*:2312:11647 (2023).

[59] Salaris, M. & Cassisi, S. Chemical element transport in stellar evolution models. *Royal Society Open Science* **4** (8), 170192 (2017).

[60] Hughto, J. *et al.* Direct molecular dynamics simulation of liquid-solid phase equilibria for a three-component plasma. *Phys. Rev. E* **86** (6), 066413 (2012).

[61] Cheng, S., Cummings, J. D., Ménard, B. & Toonen, S. Double White Dwarf Merger Products among High-mass White Dwarfs. *Astrophys. J.* **891** (2), 160 (2020).

[62] Schwab, J. Evolutionary Models for the Remnant of the Merger of Two Carbon-Oxygen Core White Dwarfs. *Astrophys. J.* **906** (1), 53 (2021).

[63] Hurley, J. R., Pols, O. R. & Tout, C. A. Comprehensive analytic formulae for stellar evolution as a function of mass and metallicity. *Mon. Not. R. Astron. Soc.* **315** (3), 543–569 (2000).

[64] Cummings, J. D., Kalirai, J. S., Tremblay, P. E., Ramirez-Ruiz, E. & Choi, J. The White Dwarf Initial-Final Mass Relation for Progenitor Stars from 0.85 to 7.5 M$_\odot$. *Astrophys. J.* **866** (1), 21 (2018).

[65] Holberg, J. B. & Bergeron, P. Calibration of Synthetic Photometry Using DA White Dwarfs. *Astron. J.* **132** (3), 1221–1233 (2006).

[66] Tremblay, P. E., Bergeron, P. & Gianninas, A. An Improved Spectroscopic Analysis of DA White Dwarfs from the Sloan Digital Sky Survey Data Release 4. *Astrophys. J.* **730** (2), 128 (2011).

[67] Dufour, P., Bergeron, P. & Fontaine, G. Detailed Spectroscopic and Photometric Analysis of DQ White Dwarfs. *Astrophys. J.* **627** (1), 404–417 (2005).

[68] Blouin, S., Dufour, P., Thibeault, C. & Allard, N. F. A New Generation of Cool White Dwarf Atmosphere Models. IV. Revisiting the Spectral Evolution of Cool White Dwarfs. *Astrophys. J.* **878** (1), 63 (2019).

[69] Gentile Fusillo, N. P. *et al.* A catalogue of white dwarfs in Gaia EDR3. *Mon. Not. R. Astron. Soc.* **508** (3), 3877–3896 (2021).

[70] Dufour, P. *et al.* The Montreal White Dwarf Database: A Tool for the Community. (eds Tremblay, P. E., Gaensicke, B. & Marsh, T.) *20th European White Dwarf Workshop*, Vol. 509 of *Astronomical Society of the Pacific Conference Series*, 3 (2017).

[71] Bildsten, L. & Hall, D. M. Gravitational Settling of $^{22}$Ne in Liquid White Dwarf Interiors. *Astrophys. J. Lett.* **549** (2), L219–L223 (2001).

[72] Binney, J., Dehnen, W. & Bertelli, G. The age of the solar neighbourhood. *Mon. Not. R. Astron. Soc.* **318** (3), 658–664 (2000).

[73] Bland-Hawthorn, J. & Gerhard, O. The Galaxy in Context: Structural, Kinematic, and Integrated Properties. *Annu. Rev. Astron. Astrophys.* **54**, 529–596 (2016).

[74] Mor, R., Robin, A. C., Figueras, F., Roca-Fàbrega, S. & Luri, X. Gaia DR2 reveals a star formation burst in the disc 2-3 Gyr ago. *Astron. Astrophys.* **624**, L1 (2019).

[75] Cassisi, S., Potekhin, A. Y., Salaris, M. & Pietrinferni, A. Electron conduction opacities at the transition between moderate and strong degeneracy: Uncertainties and impacts on stellar models. *Astron. Astrophys.* **654**, A149 (2021).

[76] Fleury, L., Caiazzo, I. & Heyl, J. The origin of ultramassive white dwarfs: hints from Gaia EDR3. *Mon. Not. R. Astron. Soc.* **520** (1), 364–374 (2023).

## Acknowledgments

We thank the referees, Leandro Althaus and two anonymous reviewers, for their valuable comments that have improved this manuscript. We are grateful to Pier-Emmanuel Tremblay for insightful discussions on the *Q*-branch problem and for a careful reading of our manuscript. We thank Ken Shen and Evan Bauer for helpful discussions that improved the work presented in this manuscript. This work has benefited from discussions at the KITP program "White Dwarfs as Probes of the Evolution of Planets, Stars, the Milky Way and the Expanding Universe'" and was supported in part by the National Science Foundation under Grant No. NSF PHY-1748958. AB is a Postdoctoral Fellow of the Natural Sciences and Engineering Research Council of Canada (NSERC) and also acknowledges support from the European Research Council (ERC) under the European Union's Horizon 2020 research and innovation programme (grant agreement no. 101002408). SB was supported by a Banting Postdoctoral Fellowship and a CITA (Canadian Institute for Theoretical Astrophysics) National Fellowship. SC acknowledges support of the Martin A. and Helen Chooljian Member Fund and the Fund for Natural Sciences at the Institute for Advanced Study, and thanks Siyu Yao for her constant encouragement and inspiration.

## Author contributions

A.B., S.B. and S.C. planned this project, each contributed new insights and research threads, and wrote the manuscript. A.B. implemented the distillation process in the STELUM code and performed the cooling calculations. S.B. provided a prescription for the implementation of distillation, performed the population synthesis simulations and subsequent data analysis, and calculated model atmospheres. S.C. provided expertise on the observational properties of the *Q* branch and identified the importance of using a thin He layer and a realistic distillation profile. S.B. provided expertise on the physics of fractionation in dense plasmas.

## Author Information

Reprints and permissions information is available at www.nature.com/reprints. The authors declare no competing financial interests. Correspondence and requests for materials should be addressed to S.B. (sblouin@uvic.ca) and A.B. (antoine.bedard@warwick.ac.uk).

## Data availability

The *Gaia* data is publicly available on the *Gaia* archive (https://gea.esac.esa.int/archive). Cooling sequences calculated for this work are provided at https://doi.org/10.5281/zenodo.10201676, as are the DQ model atmosphere synthetic magnitudes.

## Code availability

The population synthesis code is provided along with the cooling sequences at https://doi.org/10.5281/zenodo.10201676. We have opted not to make publicly available the highly specialized and multi-purpose STELUM code because of its complexity.

## Extended Data

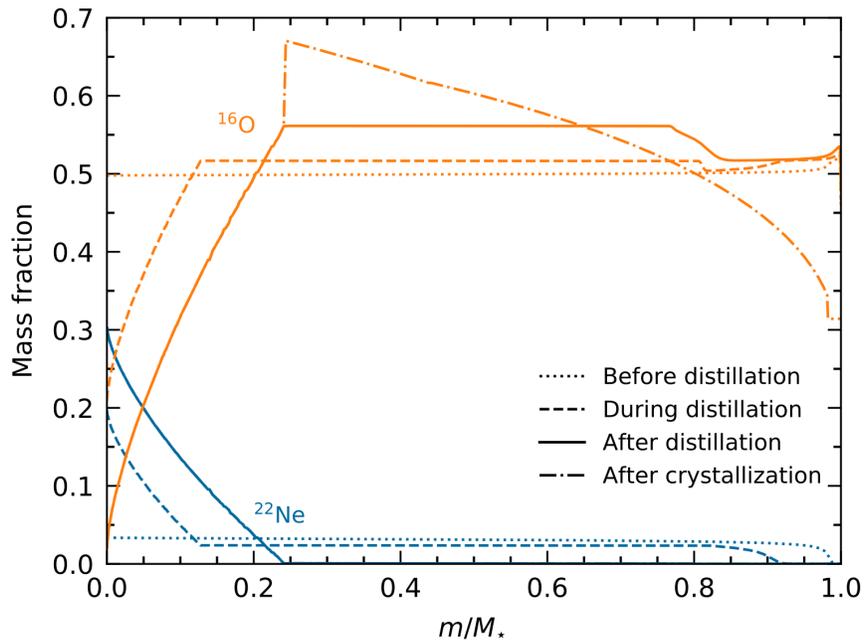

**Extended Data Figure 1 | Chemical evolution of a 1.15 $M_\odot$ white dwarf.** The $^{16}$O and $^{22}$Ne mass-fraction abundance profiles are shown at four different stages in the star's evolution.

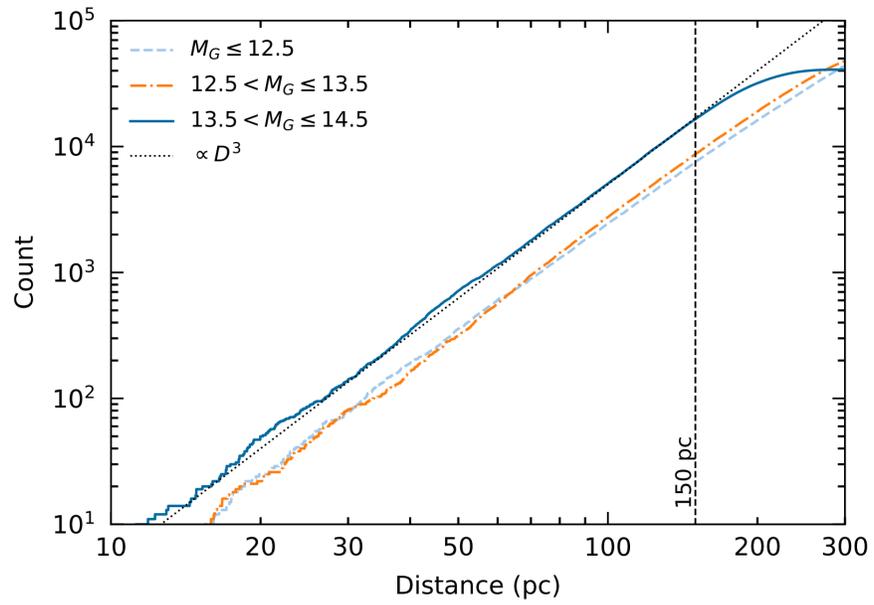

**Extended Data Figure 2 | Distance distribution of white dwarfs.** Only high-confidence white dwarf candidates ($P_{WD} \geq 0.9$) in the *Gaia* EDR3 white dwarf catalog[69] are considered. The CDF is broken down into three $M_G$ bins that span the range of magnitudes covered in our analysis of the $Q$ branch. For the faintest bin, the sample has a very high level of completeness up to a distance of 150 pc.

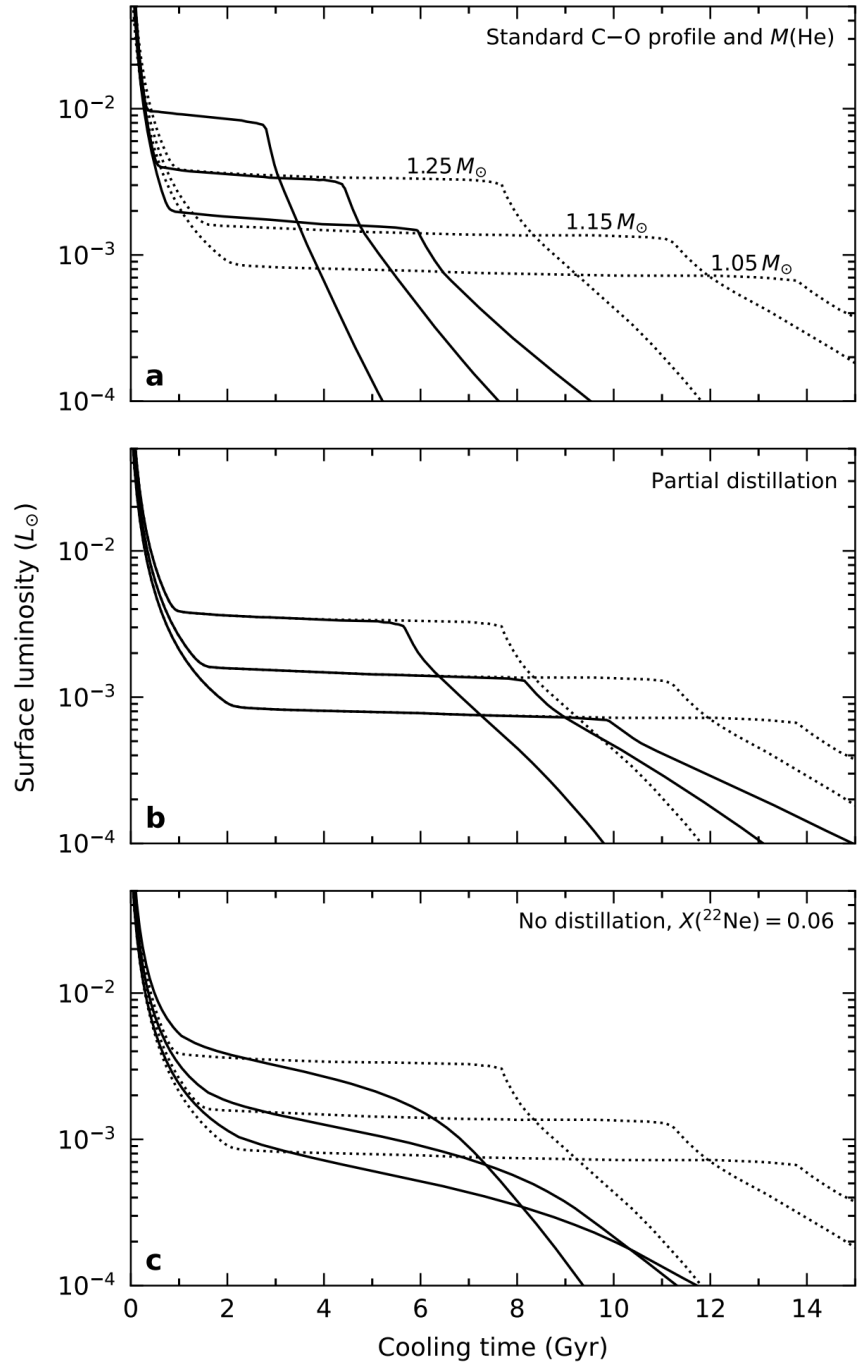

**Extended Data Figure 3 | Modeled surface luminosity for different scenarios.** The surface luminosity is shown as a function of cooling age for 1.05, 1.15, and 1.25 $M_\odot$ C-O white dwarf models. The dotted lines correspond to the nominal case considered in Fig. 4a. The solid lines represent cooling sequences where a standard composition[7] is used instead of a post-merger profile (a), where distillation is only partially completed (b), and where $X(^{22}Ne) = 0.06$ and distillation is turned off (c). Post-merger chemical profiles are assumed in panels b and c.

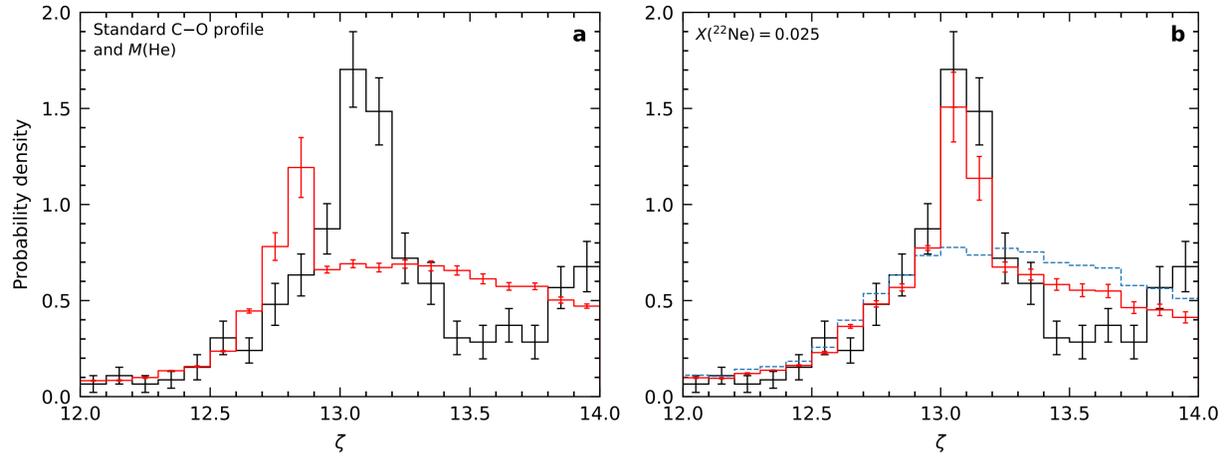

**Extended Data Figure 4 | Effect of the composition on the predicted pile-up.** Same as Fig. 3 but with different assumptions for the composition of the extra-delayed C-O population.

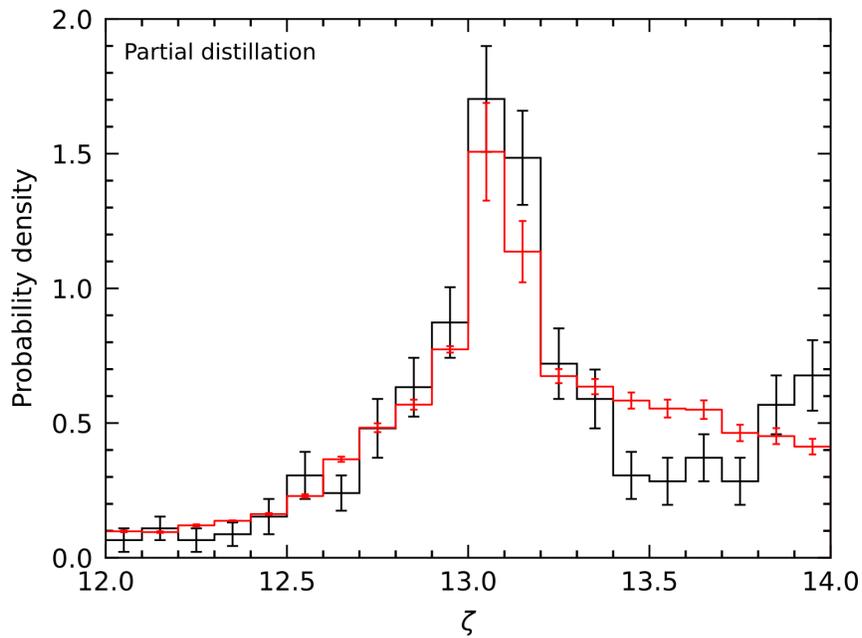

**Extended Data Figure 5 | Effect of a different implementation of distillation on the predicted pile-up.** Same as Fig. 3 but assuming partial completion of the distillation process. For the extra-delayed C-O population, our default post-merger composition profile is assumed.

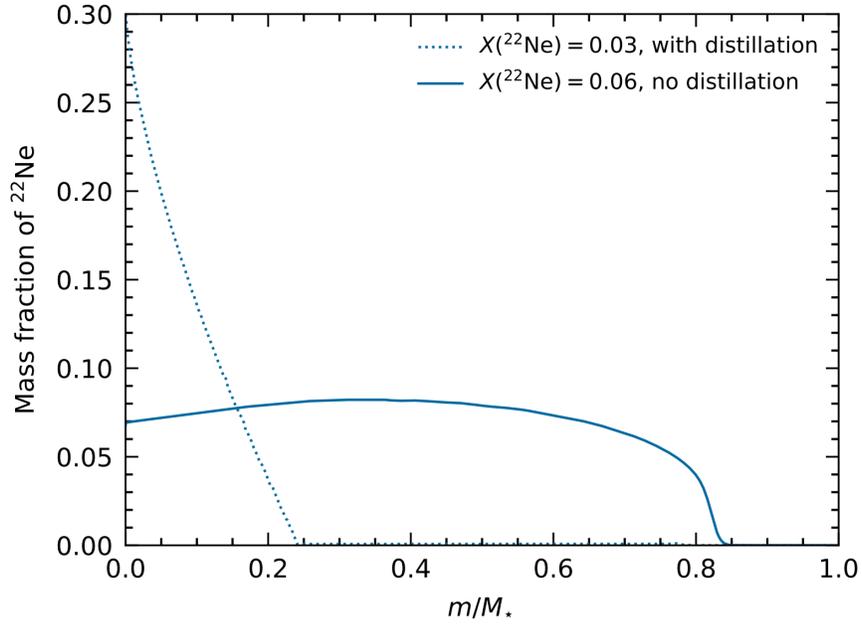

**Extended Data Figure 6 | Effect of distillation on the abundance profile.** The final $^{22}$Ne mass-fraction profile for a 1.15 M$_\odot$ white dwarf with an initial post-merger stratification is shown for two different scenarios. The composition profile is shown at the end of the crystallization process.

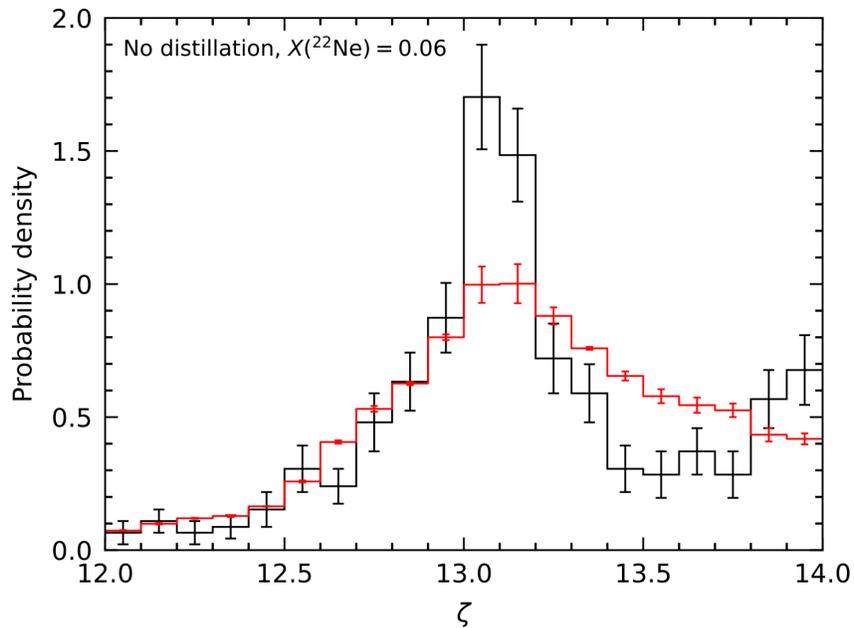

**Extended Data Figure 7 | Effect of microscopic diffusion on the predicted pile-up.** Same as Fig. 3 but with $X(^{22}\text{Ne}) = 0.06$ and no distillation (microscopic diffusion only). For the extra-delayed C-O population, our default post-merger composition profile is assumed.

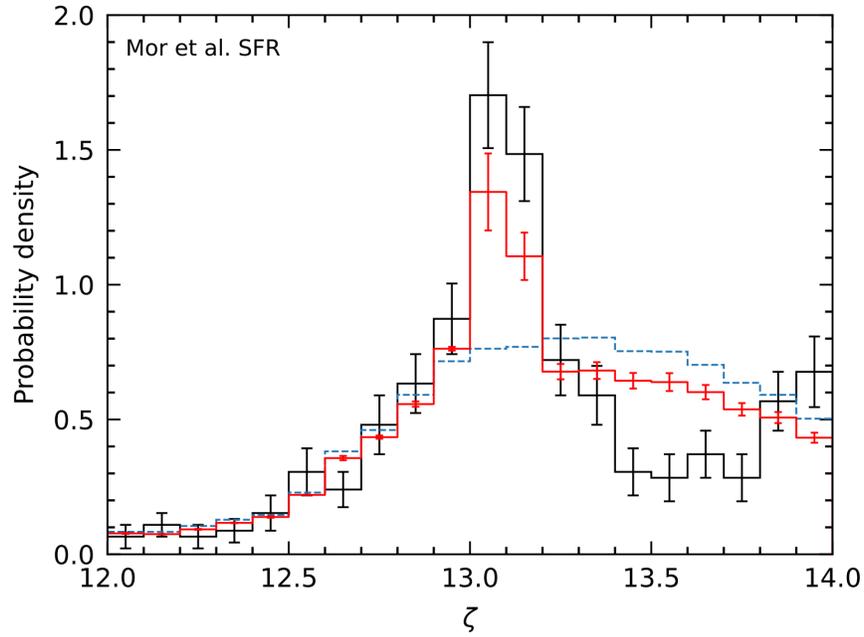

**Extended Data Figure 8 | Effect of the age distribution on the predicted pile-up.** Same as Fig. 3 but using a non-uniform stellar age distribution based on Mor *et al.*[74]

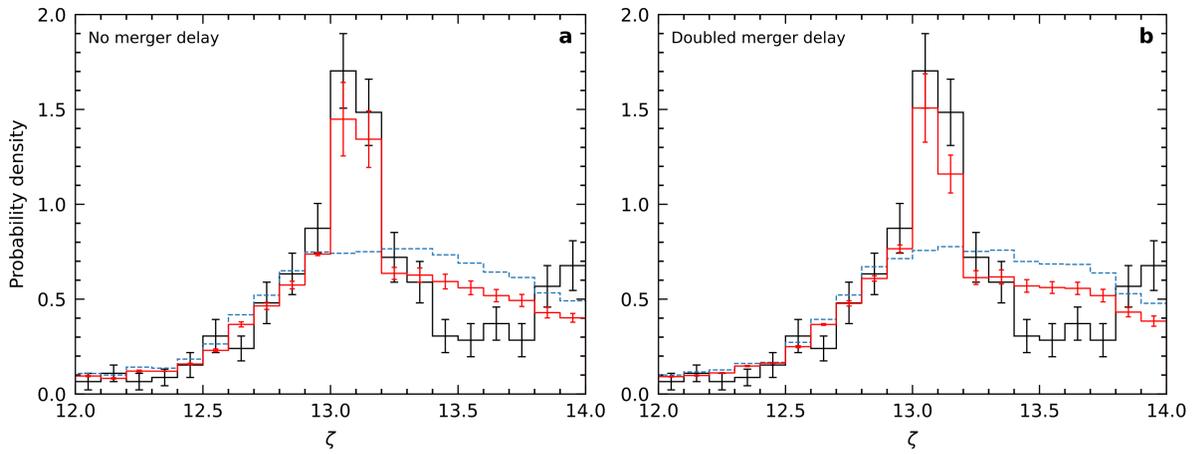

**Extended Data Figure 9 | Effect of the merger time delay on the predicted pile-up.** Same as Fig. 3 but assuming no merger time delay for all stars (a) and doubling the merger time delay used in our fiducial simulation (b).